\documentclass[opre,nonblindrev]{informs3}

\usepackage{endnotes}
\let\footnote=\endnote

\usepackage{natbib}
 \bibpunct[, ]{(}{)}{,}{a}{}{,}%
 \def\bibfont{\small}%
\usepackage{floatrow}
\usepackage[caption=false,font=footnotesize]{subfig}
\usepackage{caption}
\usepackage{csquotes}
\TheoremsNumberedThrough     
\ECRepeatTheorems

\EquationsNumberedThrough    

\begin{document}


\RUNAUTHOR{Ehsanfar and Mansouri}

\RUNTITLE{An Influence-based Clustering Model on Twitter}

\TITLE{An Influence-based Clustering Model on Twitter}

\ARTICLEAUTHORS{%
\AUTHOR{Abbas Ehsanfar}
\AFF{\EMAIL{aehsanfa@stevens.edu}} 
\AUTHOR{Mo Mansouri}
\AFF{\EMAIL{mmansour@stevens.edu}} 

\AFF{School of Systems and Enterprises, Stevens Institute of Technology, Hoboken, NJ 07030} 
} 

\ABSTRACT{%
This paper introduces a temporal framework for detecting and clustering emergent and viral topics on social networks. Endogenous and exogenous influence on developing viral content is explored using a clustering method based on the a user's behavior on social network and a dataset from Twitter API. Results are discussed by introducing metrics such as popularity, burstiness, and relevance score. The results show clear distinction in characteristics of developed content by the two classes of users. 
}%


 \HISTORY{}

\maketitle

%

\section{Introduction}

The complexity of using language on social networks in addition to active involvement of malevolent agents and bots, seizing on spreading certain topics and agenda, calls for novel explanatory approaches for detection and analysis of viral topics on social networks. In addition, the idea of formation of echo chambers among users (i.e. \textit{retweeting}) suggests that tweeters would act partially based on the sources of information~\cite{allcott2017social}, i.e. usual users act according to perceived behavior from influential users. Recent discoveries on the mechanisms of spreading fake news through influential accounts such as foreign agents and bots calls for more inclusive methods in the analysis of viral topics on a social network~\cite{ferrara2016rise,lazer2018science,vosoughi2018spread}.  

More than a decade after invention of world wide web, we started communicating through \textit{many-to-many} and interactive online social platforms. Multiple studies have developed models to understand the networked relationship among social actors and communities during this era~\cite{wang2014discovering,murase2014multilayer,sun2015social}. The networks of interactions among those are used to discover attitude, emotions, perception and sentiment associated with a content~\cite{wu2013social,west2014exploiting}. In addition, temporal interactions could reveal and predict community structure and membership behavior of users~\cite{yang2009bayesian} and to profile a user by its behavior including that of bots on social networks or agents in spreading fake news~\cite{agreste2015analysis,dickerson2014using,gilani2017bots}. In developing commercial applications, online interactions among users and analysis of topics have given insight into discovering brand reputation and political orientation~\cite{ceron2015using}. 

For analysis of topic and discourse on social networks, Davis et al. developed a ranking model for finding prevalent topics on Twitter~\cite{davis2017socirank}. Cigarr et al. presented an approach using Formal Concept Analysis (FCA) to distinguish interest groups regarding products and brands on social networks~\cite{cigarran2016step}. 
Lipizzi et. al. used a graph-based approach using adjacency matrix of concatenation among keywords to identify real-world discourses expressed through back-channeling on social networks~\cite{lipizzi2016towards} where a similar approach can cluster users based on trending topics~\cite{hachaj2017clustering}. Xie and Mathioudakis employ the concepts of popular and \textit{bursty} keywords to detect topics in real-time~\cite{xie2016topicsketch,mathioudakis2010twittermonitor}. 
Crane et al. differentiate between exogenous and endogenous topics~\cite{crane2008robust} and other studies compare topics developed on Twitter to conventional online media such as New York Times, Google trends, and CNN base on endogenous and exogenous influence on users~\cite{zhao2011comparing,kwak2010twitter}. 
Network models of users, concepts and documents have been employed to understand dynamics of community and content development on social network~\cite{gao2017network}. Network models of influential users and concepts are used to calculate the effectiveness of WikiProjects in online content development~\cite{qin2015influence} and the structure of knowledge among computer science venues~\cite{pham2010structure}. In addition, network statistics, such as centrality, closeness, betweenness, and entropy are employed to explain the interconnectedness of communities and concepts in social networks~\cite{nikolaev2015efficient,kandhway2017using}. 

This paper explores the connection among behavioral metrics of influential users and evolution of content and discourse on a social network. The scope of this paper is limited to analyze and compare two distinguished types of influential behavior in terms of their perceived contents in a network. An \textit{exogenous} influence appears when information is shared by an non-native and external source such as a news media outlet or, to some extent, a journalist. In contrary, \textit{endogenous} influence is exercised through interaction and discussion on the platform through influential users, i.e. hobs, lead users~\cite{kratzer2016social}. These observed types of behavioral influence among users are: 1) \textit{Media-driven influence (MDI)} that is associated with exogenous influence through introduction, sharing links, and summarization of topics that are originally external to a platform and 2) \textit{Interaction-driven Influence (IDI)} that is associated with interactive behaviors such as retweeting, replying, and discussing topics on the platform among others versus distributing external news and content. 

In this paper, a networked model of terms, keywords and hashtags in combination with the classified sources provide behavioral insights to temporal aspects of developing content on Twitter. This proposes a method to classify user accounts based on their online activities and distinguish two types of exogenous (MDI) and endogenous (IDI) influence which better explain the mechanism behind spreading viral topics on social networks. 
To visualize results, multiple intuitive parameters for clustering topics enhances the explanatory value of this method for our discussion regarding the nature of influence in social content. This work focuses on connecting the evolution of content on networks to behavioral characteristics of influential sources. 


\section{User Classification}
\label{sec:classification5}
In this section, we distinguish two classes of behavior by users in terms of their interaction level and type of contribution on Twitter. First, online activities of a user can be captured by her tweets, e.g. length and links, retweets, and replies. For simplicity of our behavioral model, we limit the activity model to these input metrics. This introduce a model to classify users to two classes (\textit{MDI} and \textit{IDI}) and extract new accounts on twitter associated with each class.

The introduced logistic regression model uses these features: number of retweets, number of replies, number of shared links, and median text-length:  
\begin{align}
&g(x)=\beta_0+\beta_1x_1+\beta_2x_2+\beta_3x_3+\beta_4x_4 \nonumber \\
&\pi(x)=\frac{e^{g(x)}}{1+e^{g(x)}} 
\label{eq:logistic5}
\end{align}

In the first phase, we select 170 accounts with \textit{MDI} and \textit{IDI} characteristics noticing that these accounts can be recognized by monitoring online activities of accounts in terms of sharing personal stories, direct communication with others, i.e. their organic interaction on social media. For instance, New York Times labeled as $1$ (MDI) and interactive users such as Bill Gates labeled as $0$ (IDI).
We train the logistic regression (LR) model in Eq.~\ref{eq:logistic5} using the selected accounts (training set). In second phase, we extract accounts that have been most influential on those selected accounts (e.g. being retweeted by them the most) and select second batch of data set consisting of 170 new accounts. We label the latter data set, update the logistic model and repeat the same processes of \textit{extracting}, \textit{model updating} and \textit{labeling} the most influential accounts until more than 1750 accounts are collected. In sum, $10\%$ of accounts were labeled manually and the other $90\%$ were collected and labeled iteratively (in mini-batches) using above LR model.    
\begin{table}[!h]
\centering
\caption{Features and coefficients in LR}
\label{tab:logistic5}
\begin{tabular}{|c|c|c|c|c|c|}
	\hline 
	Variable: & Interception & Retweets &  Replies & Links & Length \\ 
	\hline 
	$\beta_i$ Coefficient: & -0.96 & 0.35 & -1.76 & 2.82 & 0.61 \\ 
	\hline 

\end{tabular} 
\end{table}

\begin{figure}[t]
\centering
\includegraphics[width=0.5\linewidth]{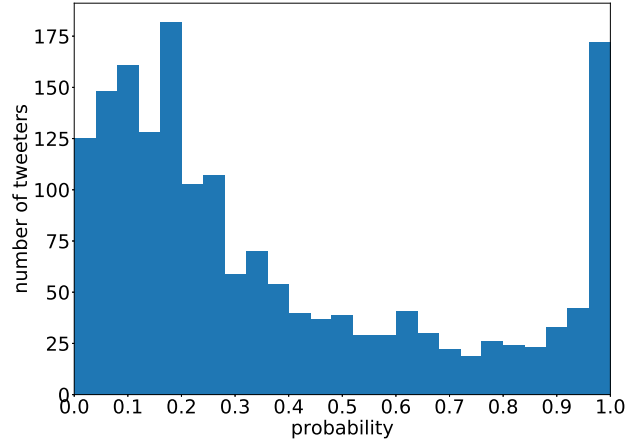}
\caption[Probability distribution after logistic regression]{\textit{Logistic regression}'s probability distribution for 1580 unlabeled samples among 1742 total accounts. The right side is associated with class $1$ or \textit{MDI} and left side relates to class $0$ or \textit{IDI}. The resulted distribution is skewed towards interactive (\textit{IDI}: left) behavior with a peak around the maximum probability. $p = 0.7$ is selected as the threshold for \textit{LR} classification.}
\label{fig:hist}
\end{figure}

Table~\ref{tab:logistic5} shows the selected variables and their calculated coefficients. The positive values for \textit{retweets}, \textit{shared links} and median \textit{length} of tweets imply that these variables are more associated with MIDI while number of \textit{replies} is more associated with IDI. 
Fig.~\ref{fig:hist} shows the probability distribution of accounts using the \textit{LR} model. 
The minimum point of $p=0.7$ is used to assign labels to accounts which results in labeling $20\%$ of users as \textit{MDI} and the rest as \textit{IDI}. 
In sum, user accounts are labeled into 352 and 1398 sources with exogenous and endogenous labels. Although the number of former accounts are significantly lower than the latter accounts, the activity level of two groups are equivalent and very close since \textit{MDI} users publish and distribute more content. 

\section{Clustering Model}
\label{sec:clustering5}

A frequentist models of words are vastly used to cluster emergence topics in social networks. In this respect, topics are both popular and scarce: first, a new topic is relatively viral and under discussion across a network, second, it may not has been as much discussed in the past or under regular circumstances. In \textit{tf-idf}, a widely used method for extracting new topics, the usage-frequency of a word in combination with the inverse-frequency of documents including the word define the relevance between the topic and a document. In social networks, a user may apply unconventional wordings, phrases, hashtags and abbreviations to efficiently communicate her message, thereby, a networked model of terms and words are applied to reconstruct grammar in analysis of discourse~\cite{himelboim2017classifying, lipizzi2016towards}. In this chapter, I also opt to employ the networked structure of language for clustering viral topics. According to a graph-based model of terms, a term $A$ (e.g. word, hashtag, abbreviation, keyword or compound word) is represented as a node when the frequency of its usage in combination with the second term $B$ in shared occasions implies the strength of connection between these terms in developing new topics across a network. 

The method introduced in this work is aimed at being efficient in detecting new topics and explanative in analysis of those topics. To discover the strength of a link between terms using term-pairs, similar to the frequency case in \textit{tf-idf}, we devise two metrics: \textit{frequency} of a pair and \textit{inverse frequency} of a pair in a time frame, i. e. lower value for a higher expected frequency. Feng et al. (see~\cite{feng2015streamcube}) defined the \textit{popularity} of an event as the normalized frequency of that event by number of tweets and \text{burstiness} of an event as the standardized popularity by the popularity of the same event achieved during other temporal time frames. 

Assume that $W_t =\{w_1, w_2, ...., w_N\}$ are the number of all words (i.e. words, hashtags, concepts, and compound words) during a time frame $t$, $D_t=\{d_{1}, d_2, ... , d_{M}\}$ are the documents (e.g. tweets, post updates) published during the same frame, and: 
$$
F_t=\{f_{ijt}: w_i,w_j \in W_t\} 
$$ 
consists of frequencies of all defined connections among those words where the cardinality of $F_t$ is the number of possible connections among $N$ words: $|F_t| = N(N-1)/2$. The \textit{popularity} of link normalizes its usage by the number of tweets during each time frame:
\begin{equation}
popularity: p_{ijt} = \frac{f_{ijt}- \mu_{t}}{\sigma_{t}}, i \in \{1, ..., L\}
\label{eq:popularity}
\end{equation}
where $\mu_{t}$ and $\sigma_{t}$ are averaged and standard deviation of frequencies in $F_t$ at time frame $t$. The latter equation which leads to an array of normalized popularity for existing links. The \textit{burstiness} standardizes popularity by the average values and standard deviation for the same connection among all temporal frames: 
\begin{equation}
burstiness: b_{ijt} = \frac{p_{ijt}-\mu_{ij}}{\sigma_{ij}}
\label{eq:burstiness}
\end{equation}
where $\mu_{ij} = \sum_{t}{p_{ijt}}/T$ is the averaged value of popularity of link between $w_i$ and $w_j$ across $T$ time frames and $\sigma_{ij}$ is the standard deviation of those frequencies. 

While popularity (Eq.~\ref{eq:popularity}) uses the direct columns of frequency matrix in time (each column represents frequencies for all connections at one time step), burstiness formula (Eq. \ref{eq:burstiness}) uses popularity matrix. A weighted average of popularity and burstiness for inter-word links gives the relevance \textit{score} for every connection: 
\begin{equation}
relevance: r_{ijt} = \alpha p_{ijt} + \beta b_{ijt}
\label{eq:relevance}
\end{equation}

A similar linear combination of above metrics in addition to \textit{localness} was called \textit{ranking score} by~\cite{feng2015streamcube}. In this work, the matrix of connection scores among terms is called as \textit{relevance matrix} which is close to definition of \textit{similarity} matrix in literature. We use a thresholds (e.g. $99^{th}$ percentile) to convert the \textit{relevance} matrix to a sparse \textit{adjacency} matrix. 


\begin{figure*}[!h]
\centering
\includegraphics[width=1\linewidth]{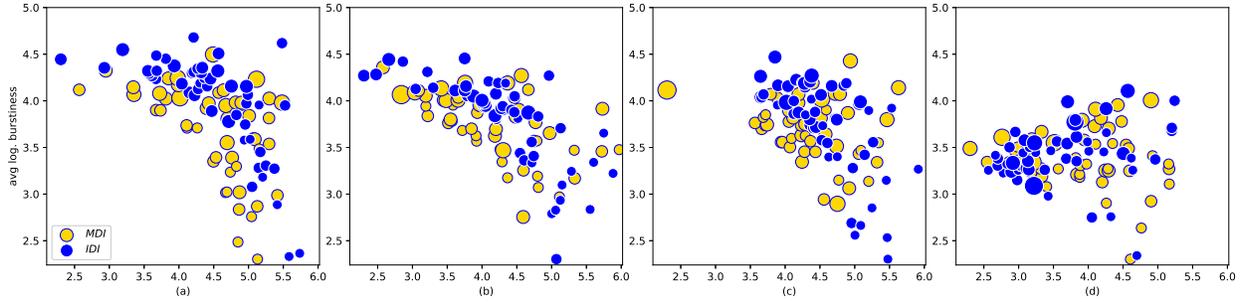}
\caption{\textit{burstiness} vs \textit{popularity} for 100 clusters of terms at each temporal granularity (50 cluster of each user class with highest \textit{cluster score}). The circle-size shows the topic consolidation in terms of relevance (see Eq.~\ref{eq:relevance}) among words in a cluster. Term-clusters among \textit{MDI} are smaller and more consistent in terms of popularity and burstiness while term-clusters resulted from tweets by \textit{IDI} accounts include relatively larger circles with more diverse sizes, which implies more distributed clusters with various strengths and consolidation for temporal granularities of: \textit{a) 1 day:}, \textit{b) 3 days:}, \textit{c) 7 days:} and \textit{d) 21 days:}. The clusters emerged from \textit{IDI} accounts in longer temporal moves towards the \textit{MDI}s as mutual effect among two classes of users emerge.}
\label{fig:popburst}
\end{figure*}

\begin{figure*}[!h]
    \centering
    \subfloat[MDI]{\label{MDImesh} \includegraphics[width=\textwidth]{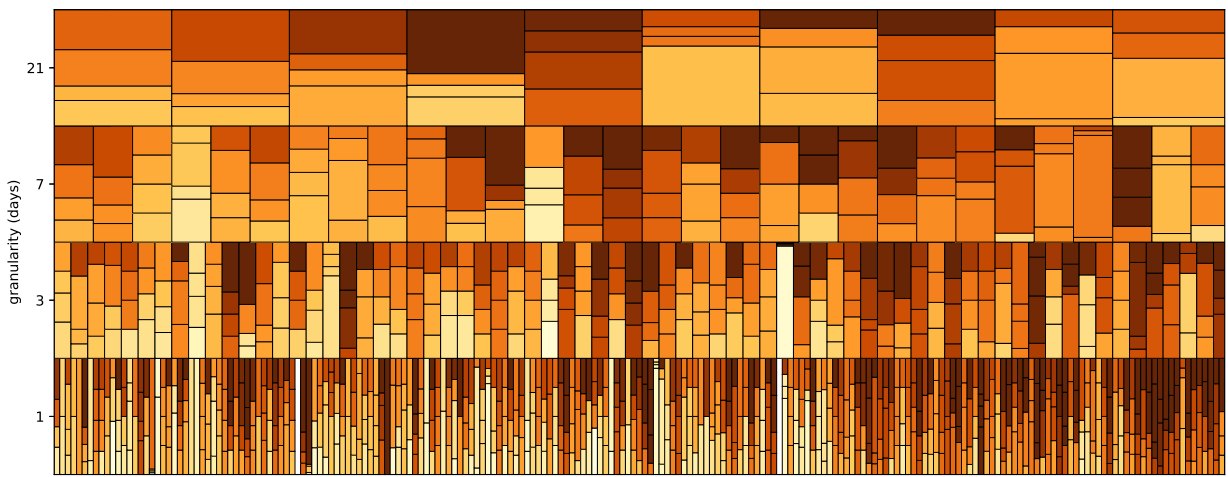}}
    \hfill
        \subfloat[IDI]{\label{IDImesh} \includegraphics[width=\textwidth]{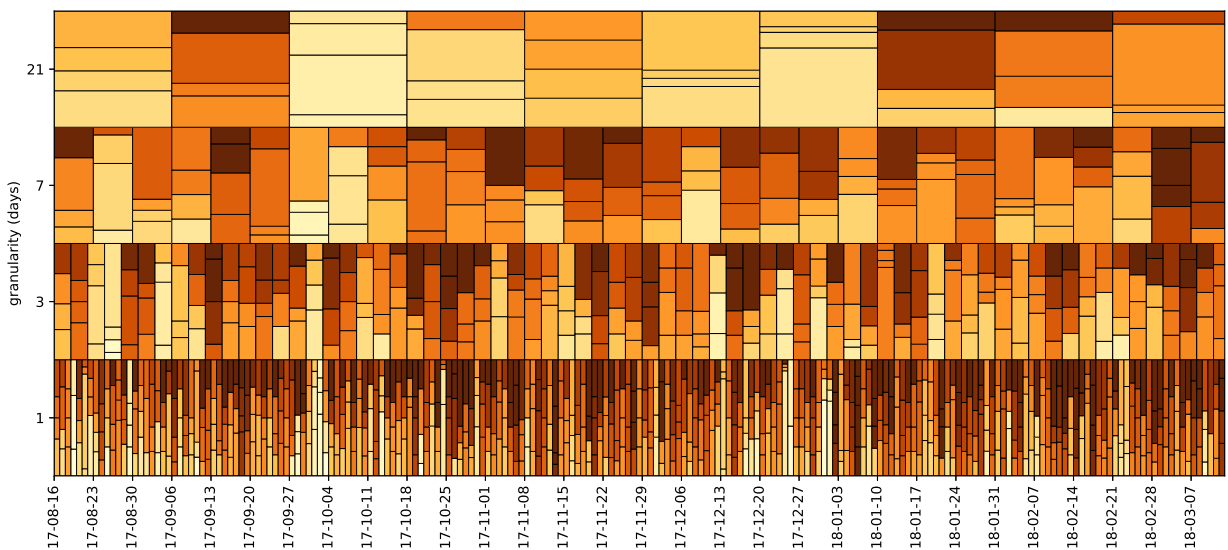}}

    \caption{dominant topic clusters for different temporal granularity: \protect\subref{MDImesh} \textnormal{\textit{MDI:} topic clusters among the most frequent terms by twitter accounts with media-driven influence,} \protect\subref{IDImesh} \textnormal{\textit{IDI:} topic clusters across network of individuals with interactive influence.} 
}
    \label{fig:clustermesh}
\end{figure*}

\section{Empirical Results and Discussion}
\label{sec:discussion5}
We collect and process more than 6,250,000 tweets published or retweeted by 1742 influential accounts on Twitter from August 2017 to March 2018. We select 355 accounts belonging to journalists, economists, scientist, news organizations, activists, etc. as initial seed of influential users\footnote{using websites such as \textit{time}, \textit{politico}, \textit{sciencemag}, etc.}. Then we retrieve another 1395 accounts from the most retweeted accounts by the initial seed during 210 days of the time period. Extracting data from Twitter API was almost sequential (10-user batch at step) to gradually update the list of most influential accounts the cumulative set of accounts determined the next batch. 

For each tweet, we stripped text from frequent words and punctuations using \textit{stopwords} repository and retrieved stemmed words using \textit{nltk} toolkit\footnote{http://www.nltk.org/howto/stem.html}. These processes significantly reduced the number of effective terms and computational complexity of graph-based model of terms in memory and time. For instance, each stem word aggregates 4.79 different words and 5000 stemmed terms are equivalent to 23950 terms. In sum, these processes reduced the number of effective statuses (by $15\%$) as it eliminated those without linguistic content e.g. photos, videos and links. For the graph-based model, we select 5000 as the maximum dimension associated with tokenizing the documents. The selected words are the words with highest frequencies across all documents. Using \textit{scipy} library on Python 3.6, we create sparse matrix of link-frequency when each link between terms $A$ and $B$ is defined as the number of tweets that include both $A$ and $B$.

\subsection{Popularity vs Burstiness}
\label{sec:popburst}
In Fig.~\ref{fig:popburst}, the relative \textit{popularity}, \textit{burstiness} for top 50 clusters in terms of their relative \textit{score} for each user-class of \textit{MDI} or \textit{IDI} are shown. In each figure, x-axis represents popularity and y-axis represents burstiness where both are relative to cluster size and logarithmic and a circle radius represents its cluster \textit{size}. The temporal granularity includes 1, 3, 7 and 21 days, from left figure to right figure. The y-axis has equal range across all temporal frames for better comparison ($2.2$ to $5$). Although these figures show the logarithmic and relative values, we use relative and non-logarithmic values of popularity, burstiness and scores for discussion.

\subsection{Viral Topics}
\label{sec:viraltopics} 

Figure~\ref{fig:clustermesh} visualizes significant clusters in terms of relative score (i.e. viralness) across time frames. We can already notice the concentration of topics for \textit{IDI} and \textit{MDI} among middle (3 and 7-day) and longer (21-day) time frames in order. In those figures, each rectangle shows a topic, darker color represents higher score and rectangle height represents cluster size. 


\section{Conclusion}
\label{sec:clustconclusion}
In this paper, we classified two sets of influential accounts based on their aggregated activity, interpreted as their endogenous and exogenous influence on the circulation of linguistic content on Twitter. 
This paper contributes: 1) a model of classifying accounts based on their activity on Twitter, 2) a clustering method for analyzing topics based on popularity and burstiness of connections among terms (term-pairs). 

The results show that the exogenous behavior of users is more limited in terms of developing diverse topics while the endogenous behavior creates greater potential to develop new topics and diversify content. After matching viral topics to actual news, the interactive behavior among users (IDI) is more focused on social, lifestyle and technology with more consistency among topics in terms of their temporal consecutiveness. On the other hand, MDI are more focused on abstract, news-based and political topics on the lowest granularity in time steps. 

Future research may develop simulation models such as Agent-based Social Systems (ABSS) based on the observations in this research to understand suspicious and adversarial behavior of influential accounts. The results from this research can be applied to design effective learning and broadcasting systems that combine diffusion of messages in social networks and interactive behavior of users.

%
%
%

\ACKNOWLEDGMENT{The authors gratefully acknowledge the preparation of Twitter data by Mr. Arash E. Vesaghi.}





\bibliographystyle{ormsv080}
\bibliography{reference.bib}
\end{document}